\documentclass[12pt,a4paper]{article}
\pdfoutput=1
\usepackage[a4paper]{geometry}

\usepackage{amsmath}
\usepackage{amsfonts}
\usepackage{amssymb}
\usepackage{dsfont}
\usepackage[nosort]{cite}
\usepackage{hyperref}
\usepackage{graphicx}
\usepackage[margin=20pt,font=small,labelfont=bf]{caption}
\usepackage[margin=0pt,font=small,labelfont=normalfont,skip=22pt]{subcaption}

\newcommand{\eq}[1]{\begin{equation}
                     \begin{split} #1 \end{split}
                     \end{equation}}
\newcommand{\ov}{\overline}
\newcommand{\op}{\hspace{1pt}}

\allowdisplaybreaks[2]
\numberwithin{equation}{section}

\begin{document}

\vspace*{1.5cm}

\begin{center}
{\LARGE
The tadpole conjecture at large \\ complex-structure
\\}
\end{center}

\vspace{0.7cm}

\begin{center}
  Erik Plauschinn
\end{center}

\vspace{0.7cm}

\begin{center} 
\textit{
Institute for Theoretical Physics, Utrecht University \\
Princetonplein 5, 3584CC Utrecht \\
The Netherlands \\
}
\end{center} 

\vspace{2cm}


\begin{abstract}
\noindent
The tadpole conjecture by 
Bena, Bl\r{a}b\"ack, Gra\~na and L\"ust
 effectively states that 
for string-theory compactifications with a large number of complex-structure moduli, 
not all of these moduli can be stabilized by fluxes.
In this note we study this conjecture
in the large complex-structure regime
using statistical data obtained by Demirtas, Long, McAllister 
and Stillman for the Kreuzer-Skarke list.
We estimate a lower bound on the flux number 
in type IIB Calabi-Yau orientifold compactifications
at large complex-structure and for large $h^{2,1}$,
and
our results support the tadpole conjecture
in this regime. 
\end{abstract}


\clearpage

\tableofcontents


\section{Introduction}

String theory is a consistent theory in ten flat space-time dimensions. Its five known 
formulations are related to each other by various dualities, and hence the theory is essentially unique. 
However, when compactifying string theory to lower dimensions one obtains
an abundance of effective theories known as the string-theory landscape. 
Famous estimates for the size of this landscape are 
$10^{500}$, $10^{930}$, $10^{1500}$ and $10^{272000}$, 
which have appeared in 
\cite{Bousso:2000xa,Schellekens:2016hhf,Lerche:1986cx,Taylor:2015xtz}.
But, although the  landscape is vastly large, 
typically not all solutions are of interest to us. For instance, one often restricts one's 
attention to effective four-dimensional theories with few or no massless 
scalar fields. 
For the type II string such theories can be obtained  by 
compactifying on Calabi-Yau three-folds in the presence of 
fluxes, where the choice of fluxes  is 
constrained by the tadpole cancellation condition
(for a review see for instance \cite{Blumenhagen:2006ci}).
Remarkably, using this tadpole condition
it has been argued that for a given compactification manifold the number of flux vacua is finite 
 \cite{Grimm:2020cda,Bakker}.

Let us become a bit more concrete:
compactifications of string theory on Calabi-Yau manifolds without fluxes 
typically lead  to a large number of massless scalar fields (moduli) 
in the effective theory. By turning on fluxes these can be stabilized  and one would 
expect that for a generic choice of fluxes all moduli can be 
stabilized in a suitable regime. This is an underlying assumption in the
KKLT \cite{Kachru:2003aw} and  
Large-Volume (LVS) \cite{Balasubramanian:2005zx} scenarios.
However, this expectation may be too naive. 
Indeed, in \cite{Bena:2018fqc} it has been argued that to 
stabilize moduli near a certain conifold locus, large fluxes have to be considered which are not 
compatible with the tadpole cancellation condition. 
In  \cite{Betzler:2019kon} we discussed that 
the tadpole cancellation condition can force moduli to be stabilized
in a perturbatively poorly-controlled regime, 
and in
\cite{Braun:2020jrx} the authors show that 
for M-theory compactifications stabilizing all moduli can be
in tension with the tadpole condition.
Thus, the landscape of four-dimensional effective theories 
with no or few massless scalar fields may be smaller than 
expected.

In \cite{Bena:2020xrh} this observation has been formulated as the \textit{tadpole conjecture}, 
and further evidence has been 
provided in \cite{Bena:2021wyr}. 
Several versions of this conjecture have been stated, and 
in this note  we are interested in the type IIB version with respect to the complex-structure 
moduli. 
With $N_{\rm flux}$ denoting the flux number encoding the contribution of 
the Neveu-Schwarz--Neveu-Schwarz (NS-NS) and Ramond-Ramond (R-R) three-form fluxes
to the D3-brane tadpole  condition, 
it states that
\eq{
  \label{tad_conj}
   N_{\rm flux}  > 2\op\alpha\op (h^{2,1}+1)
  \hspace{50pt} \mbox{for}\quad h^{2,1}\gg 1\,,
}
where 
$h^{2,1}$ counts the number of complex-structure moduli and
the constant $\alpha$ has been conjectured to be $\alpha=1/3$. 
For relevant examples  this bound is indeed satisfied (see for instance 
table~1.1 in \cite{Bena:2020xrh}).
If the tadpole  conjecture is true, then in the large $h^{2,1}$ limit 
the contribution of  fluxes to the tadpole cancellation condition cannot 
be cancelled by orientifold planes, and hence it is not possible 
to stabilize all moduli consistently. 
This would violate a common assumption of the KKLT and 
LVS constructions for large $h^{2,1}$.
We mention however that in \cite{Marchesano:2021gyv} a scenario 
evading the bound \eqref{tad_conj} has been proposed, which 
is currently under debate.

The tadpole conjecture has important implications for the string-theory landscape,
but finding a proof  appears to be difficult.
In this note we give an explanation for why it is difficult to prove this conjecture,
and we present arguments in its favor  in the large complex-structure limit. 
More concretely, 
\begin{itemize}

\item in section~\ref{sec_mod_stab} we briefly recall some aspects 
of moduli stabilization for type IIB orientifolds in the presence of fluxes. 
This section contains 
no new results and can be skipped by the reader familiar with
the topic.

\item In section~\ref{sec_tad_bound} we argue that the 
flux number $N_{\rm flux}$ appearing in \eqref{tad_conj} typically diverges
near boundaries of moduli space. This implies that 
flux-configurations with minimal $N_{\rm flux}$ will typically
stabilize moduli in the interior of moduli space
where one often has less computational control.

\item In section~\ref{sec_tad_cone} we focus on the large complex-structure limit
and determine a scaling of $N_{\rm flux}$ with $h^{2,1}$. For generic configurations we find
that in this limit $N_{\rm flux}$ exceeds the bound \eqref{tad_conj} already 
for moderate values of $h^{2,1}$, in agreement with the tadpole 
conjecture. However, configurations with smaller $N_{\rm flux}$ may
be found outside the large complex-structure region or for 
non-generic situations.

\item In section~\ref{sec_disc} we summarize and discuss the results
obtained in this work.

\end{itemize}


\section{Moduli stabilization for type IIB orientifolds}
\label{sec_mod_stab}

In this section we briefly review moduli stabilization for type IIB orientifolds 
with O3- and O7-planes. We focus on the stabilization of the 
axio-dilaton and the complex-structure moduli via 
NS-NS and R-R three-form fluxes
(for more details see for instance \cite{Grana:2005jc,Douglas:2006es}),
and the purpose of this section is to introduce the setting for this note.
However, it contains no new results and the reader familiar with the topic can  
safely skip to the next section.


\subsubsection*{Orientifold compactifications}

We  consider compactifications of type IIB string theory from ten to four dimensions 
on Calabi-Yau three-folds $\mathcal X$,
subject to an orientifold projection.
This projection contains a holomorphic involution $\sigma$ on $\mathcal X$
which  is chosen to act on the K\"ahler form and on the holomorphic three-form of $\mathcal X$ 
as $\sigma^* J= + J$ and $\sigma^*\Omega = -\Omega$. This choice gives rise 
to orientifold three- and seven-planes as its fixed-point set. 
Furthermore,  $\sigma$ splits the cohomology groups of $\mathcal X$ 
into even and odd eigen\-spaces, and relevant for our purpose is the orientifold-odd third 
cohomology of $\mathcal X$.
For this cohomology we denote an integral symplectic basis by
\eq{
\label{basis_01}
  \{ \alpha_I , \beta^I \} \in H^3_-(\mathcal X)\,, \hspace{50pt} I = 0, \ldots, h^{2,1}_- \,,
}
and for notational convenience we are going to omit the subscript of the Hodge number
in the following. 
The only non-vanishing pairings of the basis elements can be chosen as
$\int_{\mathcal X} \alpha_I \wedge \beta^J = \delta_I{}^J$, and 
we define a symplectic $(2h^{2,1}+2)\times(2h^{2,1}+2)$-dimensional matrix as
\eq{
  \label{eta}
   \eta =\int_{\mathcal X} \binom{\alpha}{\beta} \wedge  \bigl(\alpha,\beta\bigr) 
   =
   \arraycolsep3pt
   \left( \begin{array}{cc}
   0 & +\mathds 1 \\[2pt] -\mathds 1 & 0 
  \end{array}\right) .
}


\subsubsection*{Moduli}

The effective four-dimensional theory after compactification 
preserves $\mathcal N=1$ supersymmetry and 
contains  scalar fields parametrizing
deformations of $\mathcal X$.  The ones of interest to us are the axio-dilaton 
$\tau$ and the complex-structure moduli $z^i$, which we define as
\eq{
  \tau = c + i\op s\,, \hspace{80pt} z^i = u^i + i \op v^i \,, \hspace{30pt} i = 1,\ldots, h^{2,1} \,.
}
Our conventions are such that the physical domain is characterized by $s>0$ and (mostly) $v^i>0$. 
The complex-structure moduli parametrize the holomorphic three-form of the Calabi-Yau three-fold 
$\mathcal X$ as
\eq{
  \label{htf_001}
  \Omega = X^I \alpha_I - \mathcal F_I \op\beta^I \,, \hspace{60pt} z^i = \frac{X^i}{X^0} \,,
}
where the $\mathcal F_I$ depend holomorphically on the projective coordinates $X^I$. 
The K\"ahler-sector moduli $T$ and $G$  are not relevant for our discussion. 
The K\"ahler potential describing the dynamics of the moduli fields is given by 
\eq{
  \label{kpot}
  K =  - \log\bigl[ -i(\tau-\bar \tau) \bigr] - \log \left[ -i\int_{\mathcal X} \Omega \wedge \bar \Omega \right]
  - 2\log \mathcal V\,,
}
where 
the Einstein-frame volume $\mathcal V$ of the three-fold depends on the K\"ahler-sector moduli. 
Note  that we ignore $\alpha'$-corrections to the volume-term and hence the
K\"ahler potential  is of no-scale type.


\subsubsection*{Fluxes}

In order to generate a potential for the axio-dilaton and the complex-structure moduli
we consider NS-NS and R-R three-form fluxes $H_3$ and $F_3$ along the internal space
$\mathcal X$. These can be expanded into the basis \eqref{basis_01} as
\eq{
\label{tad_02}
  H_3 = h^I \alpha_I - h_I \beta^I \,, \hspace{50pt}
  F_3 = f^I \alpha_I - f_I \beta^I \,, 
}
where the expansion coefficients $h^I, h_I, f^I, f_I$ are integers due to the familiar flux-quantization 
condition. 
These fluxes generate a scalar potential in the effective four-dimensional  theory,
which is encoded in the following superpotential \cite{Gukov:1999ya}
\eq{
\label{pot_03}
 W = \int_{\mathcal X} \Omega \wedge G_3 \,, \hspace{50pt} 
 G_3 = F_3 - H_3 \op \tau \,.
}


\subsubsection*{Tadpole-cancellation condition}

Orientifold compactifications give rise to orientifold planes which are charged under the R-R gauge potentials.
They therefore contribute to the corresponding Bianchi identities as sources, and in order to solve these identities 
one typically has to introduce D-branes. For our setting these are  D3- and D7-branes. 
The integrated versions of the Bianchi identities are known as the tadpole-cancellation 
conditions, and the one relevant for us is the D3-brane tadpole given by (for  details 
on the derivation see for instance \cite{Plauschinn:2008yd})
\eq{
\label{tadpole_d3}
0=\frac{N_{\rm flux}}{2}  +  N_{{\rm D}3}  -  \frac{N_{{\rm O}3}}{4} 
  -\sum_{{\rm D}7_{\mathsf i}}\left( \frac{1}{2}\int_{\Gamma_{{\rm D}7_{\mathsf i}}}
  \hspace{-10pt}
    \mbox{tr}\left[ \mathsf F^2_{{\rm D}7_{\mathsf i}}\right] 
  + N_{{\rm D}7_{\mathsf i}}\op\frac{\chi( \Gamma_{{\rm D}7_{\mathsf i}}\bigr)}{24} 
  \right)
  -\sum_{{\rm O}7_{\mathsf j}} \frac{\chi\bigl( \Gamma_{{\rm O}7_{\mathsf j}}\bigr)}{12} \,.
}
Here,  $N_{{\rm D}7_{\mathsf i}}$ denotes the number of D7-branes in a stack labelled by $\mathsf i$
wrapping a four-cycle $\Gamma_{{\rm D}7_{\mathsf i}}$ in $\mathcal X$
and $N_{{\rm D}3}$ is the total number of D3-branes. Both of these numbers are counted 
without the orientifold images.
Furthermore, $\mathsf F_{{\rm D}7_{\mathsf i}}$ is the open-string gauge flux for a stack $\mathsf i$, 
$N_{{\rm O}3}$ is the total number of O3-planes and $\chi(\Gamma)$ denotes 
the Euler number of the cycle $\Gamma$.
The quantity which we will focus on in this paper is the flux number $N_{\rm flux}$, 
which is defined in terms of the fluxes \eqref{tad_02} as
\eq{
  \label{tad_03}
  N_{\rm flux} = \int_{\mathcal X} F_3 \wedge H_3 =  h^If_I - h_I f^I  \,.
}
(We have
chosen a convention in which  $N_{\rm flux}$ is positive in the physical domain.)


\subsubsection*{Scalar potential}

The effective four-dimensional theory resulting from compactifying type II string theory on 
Calabi-Yau orientifolds can be 
described in terms of $\mathcal N=1$ supergravity. 
The corresponding  F-term potential takes the standard form
\eq{
  \label{pot_01}
 V_F = e^K \Bigl[ F_{M\vphantom{\ov N}} G^{M\ov N} \ov F_{\ov N} - 3\lvert W\rvert^2 \Bigr] \,,
}
where $M,N$ labels the axio-dilaton, complex-structure and K\"ahler-sector moduli, 
where $F_M = \partial_M W + \partial_M K W$ denotes the F-terms and where $G^{M\ov N}$ is the inverse of 
the K\"ahler metric $G_{M\ov N} = \partial_{M\vphantom{\ov N}} \partial_{\ov N} K$.
Since the K\"ahler potential \eqref{kpot} for the K\"ahler-sector  moduli is of no-scale type and since 
$W$ is independent of the K\"ahler-sector moduli, 
\eqref{pot_01} simplifies to
\eq{
\label{pot_04}
 V_F = e^K \Bigl[ F_{A\vphantom{\ov B}}\op G^{A\ov B}\op \ov F_{\ov B}\op \Bigr]\,,
}
where $A,B$ label only the axio-dilaton and the complex-structure moduli. The global minimum of \eqref{pot_04}
therefore corresponds to vanishing F-terms, that is
$F_{A} = 0$.
As discussed  in \cite{Giddings:2001yu}, these conditions can equivalently be expressed as an imaginary
self-duality condition for $G_3$ given in \eqref{pot_03}. In particular, with $\star$ denoting the Hodge-star operator 
the F-term conditions
are equivalent to 
\eq{
  \label{eom_001}
  \star G_3 = i \op G_3 \,.
}


\subsubsection*{Hodge-star operator}

The Hodge-star operator of a Calabi-Yau three-fold appearing in \eqref{eom_001}
can be written using 
the period matrix $\mathcal N = \mathcal R + i\op \mathcal I$  defined as 
\eq{
\label{pm}
{\cal N}_{IJ}=\overline{\mathcal {F}}_{IJ}+2i \, \frac{
{\rm Im}(\mathcal F_{IM}) X^M \, {\rm Im}(\mathcal F_{JN}) X^N}{
           X^P \,{\rm Im}(\mathcal F_{PQ}) X^Q}  \,,
}
where $\mathcal F_{IJ} = \partial_I \mathcal F_J$ is symmetric in its indices 
and where 
$X^I$ and $\mathcal F_I$ are the periods introduced in \eqref{htf_001}. 
For the integral symplectic basis  $\{\alpha_I,\beta^I\}$ shown in 
\eqref{basis_01}
one can  determine the Hodge-star operator as
\eq{
\label{mat_197}
   \mathcal M = \int_{\mathcal X} \binom{\alpha}{\beta} \wedge \star \bigl(\alpha,\beta\bigr) 
  =
   \arraycolsep3pt
   \left( \begin{array}{cc}
   -\mathcal I - \mathcal R\op \mathcal I^{-1} \mathcal R
   &
  - \mathcal R\op \mathcal I^{-1} 
  \\[2pt]
  - \mathcal I^{-1}\op \mathcal R & - \mathcal I^{-1} 
  \end{array}\right) ,
}
which is a  $2(h^{2,1}+1)\times 2(h^{2,1}+1)$-dimensional, positive-definite and symmetric
matrix.


\subsubsection*{Minimum condition and flux number}

We finally want to express the imaginary self-duality condition \eqref{eom_001} in matrix formulation. 
To do so, we introduce  $2(h^{2,1}+1)$-dimensional flux vectors in the following way
\eq{
  \label{flux_850}
  \mathsf H_3 = \binom{\hphantom{-}h^I}{-h_I} \,,
  \hspace{50pt}
  \mathsf F_3 = \binom{\hphantom{-}f^I}{-f_I} \,,  
}
where we use $\mathsf H_3$ and $\mathsf F_3$ to distinguish the vectors 
\eqref{flux_850} from the differential forms $H_3$ and $F_3$ given in \eqref{tad_02}. 
The minimum condition  \eqref{eom_001} can then be written 
using the symplectic pairing $\eta$ shown in \eqref{eta} as 
\eq{
\label{eom_104}
\mathsf F_3 = \rho \,\mathsf H_3 \,,
\hspace{70pt} \rho = \eta\op \mathcal M \op s +  \mathds 1 \op c\,,
}
and the eigenvalues of $\rho$ are $c\pm i\op s$.
We also note that using this relation, the flux number $N_{\rm flux}$ 
(evaluated for a solution of \eqref{eom_104})
can be expressed in the following three equivalent ways
\eq{
\label{tadpole_740}
  N_{\rm flux} 
\hspace{5pt}=\hspace{5pt} \mathsf F_3^T\op \eta\, \mathsf H_3^{\vphantom{T}} 
\hspace{5pt} =\hspace{5pt} s\,\bigl(\op \mathsf H_3^T \mathcal M\, \mathsf H_3^{\vphantom{T}} \bigr)
\hspace{5pt}  = \hspace{5pt} \frac{s}{c^2 + s^2}\, \bigl( \op\mathsf F_3^T \mathcal M \,\mathsf F_3^{\vphantom{T}} \bigr)\,.
}


\vskip1em
\section{Boundary behavior of $N_{\rm flux}$}
\label{sec_tad_bound}

In this section we study the behavior of the flux number $N_{\rm flux}$ 
when approaching a boundary in complex-structure or 
axio-dilaton moduli space. 
We argue that in such a limit the flux number typically diverges,
which for  the type IIB $\mathbb T^6/\mathbb Z_2\times \mathbb Z_2$ orientifold
has been observed already in \cite{Betzler:2019kon}
and which in the context of asymptotic Hodge theory has been shown in 
\cite{Grimm:2020cda}.
We note however that our arguments do not constitute a 
proof of the above statement but rather illustrate a generic behavior.


\subsubsection*{Bloch-Messiah decomposition}

For ease of presentation, let us first define $n=h^{2,1}+1$. 
The matrix $\mathcal M$ representing the Hodge-star operator 
 shown in \eqref{mat_197}  is a 
real, symmetric, symplectic  matrix. Indeed, given the explicit form 
of $\mathcal M$  and using the symplectic pairing \eqref{eta}
one can verify that
\eq{
 \mathcal M^T \eta \, \mathcal M  = \eta \,,
}
and hence $\mathcal M \in \mathrm{Sp}(2n,\mathbb R)$. 
We can then perform a Bloch-Messiah decomposition of $\mathcal M$ as
\eq{
  \label{decomp_402}
  \mathcal M = U^T \op \Sigma \, U \,,\hspace{40pt}
  \Sigma = \left( \begin{array}{cc} \lambda & 0 \\ 0 & \lambda^{-1} \end{array}\right), 
  \hspace{10pt} 
  U \in \mathrm{Sp}\bigl(2n,\mathbb R\bigr) \cap \mathrm{O}(2n,\mathbb R) \,,
}
where $\lambda$ is a $n$-dimensional diagonal matrix with entries $\lambda_I$ 
and $U$ is a symplectic ($U^T\op\eta \, U = \eta$) 
and orthogonal ($U^T U = \mathds 1$) matrix. 
We require $\mathcal M$ to be strictly positive-definite inside the 
complex-structure moduli space, so it follows that the eigenvalues $\lambda_I$ and $\lambda_I^{-1}$  appearing in 
$\Sigma$ are all positive in this region. 
However, at a boundary of the moduli space the matrix $\mathcal M$ 
is expected to degenerate. Since $U$ is non-singular this implies that at least one eigenvalue of $\mathcal M$ 
has to vanish and at least one eigenvalue has to diverge.


\subsubsection*{Minimum conditions}

Let us now turn to the minimum condition \eqref{eom_104}. Using that the matrix $U$ in 
\eqref{decomp_402} is symplectic and orthogonal, we can bring \eqref{eom_104} into  the 
following form
\eq{
  \label{eom_621}
  \widetilde{\mathsf F}_3 = 
 \left( \begin{array}{cc} \mathds 1\op c &  \lambda^{-1} \op s  \\ - \lambda\op s & \mathds 1\op c \end{array}\right)
 \widetilde{\mathsf  H}_3   \,,
  \hspace{50pt}
  \widetilde{\mathsf H}_3 = U\op \mathsf H_3 \,,\quad
  \widetilde{\mathsf F}_3 = U\op \mathsf F_3 \,,
}
where matrix notation is again understood.
Note that each of the four blocks of the matrix in \eqref{eom_621} is a diagonal matrix. 
Let us now focus on a single index-combination in \eqref{eom_621}, which without loss of generality
we choose as 
\eq{
  \label{eom_963}
  \arraycolsep2pt
  \begin{array}{lcl}
  \tilde{f}^0 &=& c\,  \tilde{h}^0 + \lambda_0^{-1} \op s \op  \tilde{ h}_0 \,,
  \\[4pt]
  \tilde{f}_0 &=& c\,  \tilde{ h}_0 - \lambda_0^{\hphantom{-1}} \op s \op  \tilde{h}^0 \,.
  \end{array}
}
Similarly as in \eqref{flux_850} $\tilde h^I$, $\tilde h_I$ and $\tilde f^I$, $\tilde f_I$ are
the components of $\tilde{\mathsf H}_3$ and $\tilde{\mathsf F}_3$, which 
are now however
neither constant nor integer-valued. 
As can be inferred from \eqref{eom_963}, in order to stabilize the (combination of moduli parametrizing the) eigenvalue 
$\lambda_0$,
the fluxes $\tilde{h}^0$ and $\tilde{f}^0$ cannot both be zero. 
Using that the matrix $U$ is symplectic, 
we find from \eqref{tadpole_740} that the flux number takes the form
\eq{
  \label{tadpole_951}
  \arraycolsep2pt
  \begin{array}{lcl@{\hspace{50pt}}lcclclll}
  \tilde{ h}^0 &\neq &0 &
  N_{\rm flux} &=&\displaystyle  s& \displaystyle \Bigl[ \op\lambda_0\op \bigl(\tilde{ h}^0\bigr)^2 &+& \displaystyle
  \tfrac{1}{\lambda_0} \op \bigl(\tilde{ h}_0\bigr)^2 &+&
   \ldots \Bigr]\,,
  \\[8pt]
  \tilde{ f}^0 &\neq &0 &
  N_{\rm flux} &=&\displaystyle  \frac{s}{c^2 + s^2}& \displaystyle \Bigl[ \op\lambda_0\op \bigl(\tilde{ f}^0\bigr)^2 
  &+& \displaystyle
  \tfrac{1}{\lambda_0} \op \bigl(\tilde{ f}_0\bigr)^2 &+&
   \ldots \Bigr]\,,
  \end{array}
}
where the ellipses in the parenthesis denote additional semi-positive terms for the remaining index-pairs.
Of course, in the case that $\tilde{ h}^0$ and $\tilde{ f}^0$ are both non-zero, 
the two expressions in \eqref{tadpole_951} are equal to each other.


\subsubsection*{Boundary limit I}

We now want to study the flux number $N_{\rm flux}$ in a particular boundary limit. 
As mentioned 
above, at a boundary in complex-structure moduli space the Hodge-star matrix 
$\mathcal M$ is expected to degenerate and  we parametrize this degeneration without 
loss of generality by 
\eq{
  \label{exp_0010}
  \lambda_0 \to \infty\,.
}
Scaling an eigenvalue of a matrix does not change its eigenvector, and hence
the matrix $U$ introduced in 
\eqref{decomp_402} does not change under \eqref{exp_0010}.
Since the fluxes $\mathsf H_3$ and $\mathsf F_3$ defined in \eqref{flux_850}
are integer quantized, it follows that in the above limit \label{exp_001}
$\widetilde{\mathsf H}_3 = U\op \mathsf H_3$ and 
$\widetilde{\mathsf F}_3 = U\op \mathsf F_3$ 
cannot be made arbitrarily small by a choice of fluxes.
In particular, a non-vanishing $\tilde{\mathsf h}^0$ or $\tilde{\mathsf f}^0$
has a lowest allowed value. 
It therefore follows that 
\eq{
    \lambda_0 \to \infty \hspace{40pt}\Longrightarrow \hspace{40pt}
    N_{\rm flux} \to  \infty\,,
}
and hence the flux number diverges when approaching a boundary of 
complex-structure moduli space. 
Let us emphasize that  the main assumption for this argument is that the 
boundary limit \eqref{exp_0010} 
can indeed be realized for a concrete moduli-space geometry with, in particular, the 
matrix $U$ remaining invariant or changing only slightly. This assumption may not be realized
in specific models.


\subsubsection*{Boundary limit II}

Let us now discuss a different limit: the Hodge-star matrix \eqref{mat_197} 
depends on $2\op h^{2,1}$ real moduli fields, whereas the 
eigenvalue matrix $\Sigma$ shown in \eqref{decomp_402} has
at most $h^{2,1}+1$ independent degrees of freedom. 
Therefore, the  matrix $U$ has a moduli dependence which 
could in principle compensate the growth in $\lambda_I$. 
To illustrate this possibility we  consider the case 
$\tilde{ h}^0\neq0$. If we require that $N_{\rm flux}$ in \eqref{tadpole_951}
stays finite in the limit $\lambda_0\to\infty$,
$\tilde{ h}^0$ and $\tilde f_0$ have to scale as 
\eq{
  \tilde{ h}^0 \sim (\lambda_0)^{-\frac{1}{2}}
  \qquad\xrightarrow{\quad \eqref{eom_963} \quad}\qquad
  \tilde{ f}_0 \sim (\lambda_0)^{+\frac{1}{2}} \,.
}
(For $c\neq0$ one can also choose $ \tilde{ h}_0 \sim (\lambda_0)^{+\frac{1}{2}} $
and $\tilde f_0$ finite, 
for which the following argument goes through analogously.)
The matrix $U$ appearing in \eqref{eom_621} is orthogonal, and so we have the relation
\eq{
  \label{tad_8883}
  \lambda_0 \sim   (\tilde{ f}_0)^2 \leq \lVert \op\widetilde{\mathsf F}_3 \rVert^2
  =  \lVert \op\mathsf F_3 \rVert^2 
  \hspace{20pt}\Longrightarrow\hspace{20pt}
  \lVert \op\mathsf F_3 \rVert^2 
  \xrightarrow{\hspace{6pt}\lambda_0\to \infty\hspace{6pt}} \infty \,,
}
where $\lVert\op \cdot\op\rVert$ denotes the ordinary Euclidean norm. 
Hence, if we require the flux number $N_{\rm flux}$ to stay finite when approaching 
a boundary, the sum of squared $F_3$-flux quanta 
has to diverge.\footnote{When relaxing the requirement that 
the contribution to the flux number stays finite  and allowing the first two terms in 
the parenthesis of \eqref{tadpole_951} to go to zero, the modulus 
associated to the eigenvalue $\lambda_0$ becomes massless. 
We exclude such a behavior.}
A similar conclusion is reached for the case $\tilde{ f}^0\neq0$. 
Within the framework used in this section we cannot exclude such a mechanism, 
although we consider it non-trivial  that it can be realized in concrete examples.
Certainly, such flux choices are non-generic.


\subsubsection*{Boundary limit III}

Turning finally to the axio-dilaton $\tau=c+i\op s$, we note that we can restrict its moduli space to the region 
$c^2 + s^2 \geq 1$, $\lvert c\rvert \leq \tfrac{1}{2}$ using S-duality. The boundary then corresponds to $s\to\infty$, 
and we see from 
$N_{\rm flux} =  s\,( \mathsf H_3^T \mathcal M\, \mathsf H_3^{\vphantom{T}})$
shown in \eqref{tadpole_740} that 
\eq{
    s \to \infty \hspace{40pt}\Longrightarrow \hspace{40pt}
    N_{\rm flux} \to  \infty\,,
}
provided one keeps the complex-structure moduli fixed, i.e.~they remain stabilized 
in a localized region of moduli space when taking $s\to\infty$.


\subsubsection*{Summary and remarks}

We close this section with the following summary and remarks:
\begin{itemize}

\item We have presented arguments that when approaching a boundary in complex-structure 
or axio-dilaton moduli space, the flux number $N_{\rm flux}$ diverges. 
This implies that in order to obtain a minimal $N_{\rm flux}$ one has to stabilize moduli in the 
interior of moduli space, sufficiently-far away from any boundaries.
However, in such regions one  has less computational control over 
the moduli-space geometry and 
it can  become difficult to find well-controlled potential
counter-examples to the tadpole conjecture \eqref{tad_conj}.

\item The above conclusion applies to generic situations, and
a possible loophole is to choose specific fluxes with large Euclidean norms
(c.f.~equation \eqref{tad_8883}) that nevertheless lead to a small 
flux number $N_{\rm flux}$.  
We believe that such directions in flux-space are non-generic. Furthermore, 
we note that especially for large $h^{2,1}$ it is unlikely 
to find them using Monte-Carlo methods.

\item The observation in this section can also be made using asymptotic Hodge theory. 
Here one can choose coordinates such that 
a boundary in complex-structure moduli space corresponds to 
the limit $\mbox{Im}\, z = v \to \infty$, and one finds that when approaching 
this  boundary for a single modulus the flux number diverges \cite{Schmid,CKS,Grimm:2020cda}, that is
\eq{
  \label{tad_9375}
  v\to \infty \hspace{40pt}\Longrightarrow \hspace{40pt}
    N_{\rm flux} \to  \infty\,.
}

\end{itemize}


\section{Tadpole conjecture at large complex-structure}
\label{sec_tad_cone}

We now want to study the tadpole conjecture in the large complex-structure 
limit. In section~\ref{sec_tad_lcs} we determine
the moduli dependence of the flux number $N_{\rm flux}$,
in section~\ref{sec_prop_cone} we review results of \cite{Demirtas:2018akl} on 
how for large values of $h^{2,1}$ the ``complex-structure cone''  becomes 
narrow, and in section~\ref{sec_scaling} we estimate the dependence of $N_{\rm flux}$ 
on $h^{2,1}$ for generic flux compactifications and compare it with the tadpole conjecture.


\subsection{The large complex-structure regime}
\label{sec_tad_lcs}

We start by briefly recalling some properties of the complex-structure moduli 
space in the large complex-structure regime,
and we determine the scaling behavior of the flux number $N_{\rm flux}$ in 
this limit.


\subsubsection*{Large complex-structure limit}

We consider a Calabi-Yau three-fold $\mathcal X$ in the large complex-structure limit, 
for which the moduli-space
geometry can be described using K\"ahler-moduli data of the mirror-dual manifold. 
The corresponding prepotential takes the form
\eq{
\label{prepot_01}
  \mathcal F = -\frac{1}{3!}\op \frac{\kappa_{ijk}\op X^i X^j X^k}{X^0}  \,,
}
where $\kappa_{ijk}$ are the triple intersection numbers of the mirror-dual three-fold $\mathcal Y$.
The $X^I$ are the projective coordinates appearing in the holomorphic three-form \eqref{htf_001} 
and the periods $\mathcal F_I$ of $\Omega$ are computed from \eqref{prepot_01} 
as $\mathcal F_I = \partial_I \mathcal F$. 
For this prepotential, the real and imaginary parts of $\mathcal N$ defined in \eqref{pm} can 
be determined as follows 
\eq{
\label{periodlargecc}
  \arraycolsep2pt
  \renewcommand{\arraystretch}{1.35}
  \begin{array}{lcl@{\hspace{50pt}}lcl}
  {\rm Im}\, \mathcal{N}_{ij} &=& -\frac{2}{3} \op\kappa\, G_{i\ov j} \,,&
  {\rm Re}\, \mathcal{N}_{ij} &=& -\kappa_{ijk}\, u^k \,,  
  \\
  {\rm Im}\, \mathcal{N}_{i0} &=& +\frac{2}{3}\op\kappa\, G_{i\ov j}\, u^j \,, &
  {\rm Re}\, \mathcal{N}_{i0} &=& +\frac{1}{2}\op\kappa_{ijk}\, u^j u^k \,,
  \\
  {\rm Im}\, \mathcal{N}_{00} &=&  - \frac{2}{3}\op\kappa \, G_{i\ov j}\, u^i u^j  -\frac{1}{6}\op \kappa\,, &
  {\rm Re}\, \mathcal{N}_{00} &=& -\frac{1}{3}\op \kappa_{ijk}\, u^iu^j u^k \,,
  \end{array}
}  
where we recall that $z^i = u^i + i \op v^i$ and where we use the notation
$ \kappa = \kappa_{ijk} v^i v^j v^k$, $\kappa_i = \kappa_{ijk} v^j v^k $ and $ \kappa_{ij} = \kappa_{ijk}  v^k$.
The K\"ahler metric $G_{i\ov j}$ follows from the K\"ahler potential \eqref{kpot} as
\eq{
\label{geo_01}
G_{i\ov j} = -\frac{3}{2}\op \frac{\kappa_{ij}}{\kappa} + \frac{9}{4} \op\frac{\kappa_i \kappa_j}{\kappa^2} \,.
}
Using these expressions  and defining $\mathsf f^I = f^I - c \op h^I$ and 
$\mathsf f_I = f_I - c \op h_I$ for notational convenience, 
the minimum conditions \eqref{eom_104} can be brought into the following form\footnote{In all examples we
have studied, we noticed that
for $h^0=f^0=0$ it is not possible to stabilize all complex-structure and axio-dilaton moduli. However, we 
have no formal proof of this observation.\label{foot_modstab}}
\eq{
  \label{eom_012}
  &
  \arraycolsep2pt
  \begin{array}{lclcc@{\hspace{1pt}}lclclcl@{\hspace{1pt}}l}
  0 &=& \frac{\kappa}{6} \op \mathsf f^0  &+& \bigl( &h_0 &+&  h_i \op u^i &- &\mathcal R_{0I} \op h^I  
  &-& u^i\op \mathcal R_{iJ} \op h^J &\bigr) \op s\,, 
  \\[4pt]
   0 &=& \frac{\kappa}{6} \op h^0 \op s &-& \bigl( &\mathsf f_0 &+&\mathsf f_i \op u^i &-&\mathcal R_{0I}\op  \mathsf f^I 
  &-& u^i \op\mathcal R_{iJ} \op \mathsf f^J& \bigr) \,, 
  \end{array}
  \\[2pt]
  &
  \arraycolsep2pt
  \begin{array}{lcl@{\hspace{1pt}}l@{\hspace{1pt}}lclcl@{\hspace{1pt}}l}
  0&=& \frac{2}{3}\op\kappa\op G_{i\ov j} & ( \mathsf f^j - u^j\op \mathsf f^0 &) &+& (\op h_i &-& \mathcal R_{iJ} \op h^J &) \op s\,,
   \\[4pt]
  0 &=& \frac{2}{3}\op\kappa \op G_{i\ov j} & (h^j - u^j\op h^0 &) s &-& (\op \mathsf f_i  &-& \mathcal R_{iJ} \op \mathsf f^J&) \,.
  \end{array}
}
With the help of these relations the flux number $N_{\rm flux}$ at the minimum
can be rewritten in the following way
\eq{
\label{tad_04}
 N_{\rm flux} =
 \arraycolsep1pt
 \begin{array}[t]{rlllllllllllll}
   \displaystyle s &\Bigl[ & \frac{\kappa}{6} & (h^0)^2 &+& \frac{2}{3}\op\kappa& (h^i  &-& u^i h^0) & G_{ij} & (h^j  &-& u^j h^0) &\Bigr]
  \\[8pt]
  + \op\displaystyle \frac{1}{s} &\Bigl[ & \frac{\kappa}{6} & (\mathsf f^0)^2 &+& \frac{2}{3}\op\kappa& (\mathsf f^i  &-& u^i \mathsf f^0) 
  & G_{ij} & (\mathsf f^j  &-& u^j \mathsf f^0) &\Bigr]\,.
  \end{array}
}
Note that the metric $G_{i\ov j}$ is positive definite, that
$\kappa$ is required to be positive  and that in the physical regime $s>0$.
Hence, $N_{\rm flux}$ is a sum of semi-positive definite terms.


\subsubsection*{Scaling behavior of $N_{\rm flux}$}

We now want to estimate the behavior of $N_{\rm flux}$ when approaching the 
large com\-plex-structure point while keeping the dilaton $s$ finite. 
To do so, we first introduce a normalized vector $\mathsf v^i$ via
\eq{
  v^i =  \lVert v \rVert\, \mathsf v^i \,, \hspace{70pt} \lVert \mathsf v \rVert =1\,,
}
where  $\lVert\op \cdot\op\rVert$ denotes  the standard Euclidean norm. 
We then recall that  in \cite{Demirtas:2018akl} the authors determined properties of the 
triple-intersection numbers $\kappa_{ijk}$ of the mirror-dual three-fold 
for the Kreuzer-Skarke database \cite{Kreuzer:2000xy}. We use this information to make the following estimates:
\begin{itemize}

\item We start by considering the first terms in \eqref{tad_04} containing $h^0$ and $\mathsf f^0$. 
In figure~1 of  \cite{Demirtas:2018akl} it is shown
how the number of non-zero entries of the triple-intersection numbers depends 
on the dimension of the moduli space. From this data we can extract the following lower bound\footnote{
We emphasize that in the basis used in \cite{Demirtas:2018akl}, the K\"ahler cone is 
parametrized not necessarily by $v^i>0$ but by linear relations imposed on the $v^i$. We use this basis also in this work.}
\eq{
  \label{prop_61209}
  \#(\kappa_{ijk}\neq 0) \gtrsim 6.5\op h^{2,1} + 25
  \hspace{40pt}\mbox{for}\hspace{10pt} h^{2,1} \gtrsim 25 \,.
}
Furthermore, the average value of the components of the normalized vector $\mathsf v^i$ scales as
$\mathsf v^i \simeq (h^{2,1})^{-1/2}$, and together with 
\eqref{prop_61209} we can then estimate that
$\kappa_{ijk} \mathsf v^i \mathsf v^j \mathsf v^k \gtrsim 6.5 \op (h^{2,1})^{-1/2}$.
For large $h^{2,1}$ the expression $ \kappa = \kappa_{ijk} v^i v^j v^k$ therefore behaves as
\eq{
  \label{scaling_kappa}
  \frac{\kappa}{6}\gtrsim  (h^{2,1})^{-1/2} \op\lVert v \rVert^3 \,.
}

\item Next, we consider the remaining terms in \eqref{tad_04} proportional to
$\tfrac{2}{3}\op\kappa\op G_{i\ov j}$. 
The explicit form of $G_{i\ov j}$ was given in \eqref{geo_01}, and 
using \eqref{prop_61209} we estimate that generically 
$\kappa_i$ scales as
$\kappa_i \simeq (h^{2,1})^{-1} \lVert v \rVert^2$
while $\kappa_{ij}$ is mostly diagonal and behaves as
$\kappa_{ii} \simeq (h^{2,1})^{-1/2} \lVert v \rVert$.
Furthermore, $\tfrac{2}{3}\op\kappa\op G_{i\ov j}$
is contracted with flux vectors, which generically have 
$h^{2,1}$ non-vanishing components. 
Combining these estimates
we then arrive at
\eq{
  \tfrac{2}{3} \op\kappa\op G_{i\ov j}\, (\mbox{flux})^i\,(\mbox{flux})^j \gtrsim  (h^{2,1})^{+1/2} \op\lVert v \rVert \,,
}
where $(\mbox{flux})^i$ stands for the components of a generic flux-vector appearing in \eqref{tad_04}.

\end{itemize}
Coming back to the flux number \eqref{tad_04}, 
for finite values of $s$
we can now estimate the following behavior
in the large complex-structure regime ($\lVert v \rVert \gg 1$) and for large $h^{2,1}$ 
\eq{
\label{scaling_838}
   N_{\rm flux} \gtrsim \left\{
  \begin{array}{l@{\hspace{15pt}}l}
   \displaystyle (h^{2,1})^{+1/2} \,\lVert v\rVert & \mbox{for $h^0=f^0=0$}\,,
   \\[6pt]
   \displaystyle (h^{2,1})^{-1/2} \, \lVert v\rVert^3 & \mbox{else} \,.
   \end{array}
   \right.
}
In view of our observation in footnote~\ref{foot_modstab}, we remark that in order to 
stabilize all complex-structure and axio-dilaton moduli it appears that $h^0$ and $f^0$ cannot both be zero.


\subsection{Properties of the mirror-dual K\"ahler cone}
\label{sec_prop_cone}

In this subsection we review some additional results of \cite{Demirtas:2018akl} on how 
on the mirror-dual side
for large values of 
$h^{1,1}$ the K\"ahler cone for the K\"ahler moduli tends to become narrow.
Via  mirror symmetry, this implies for our situation that the corresponding ``complex-structure cone'' becomes
narrow for large $h^{2,1}$. 


\subsubsection*{(Stretched) K\"ahler cone}

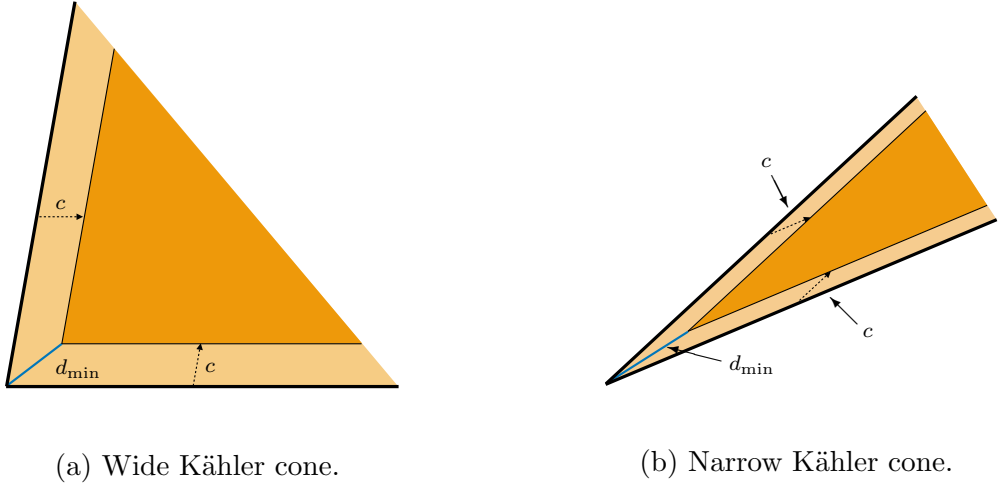
\begin{figure}
\centering
\begin{subfigure}{0.35\textwidth}
\centering
\includegraphics[width=150pt]{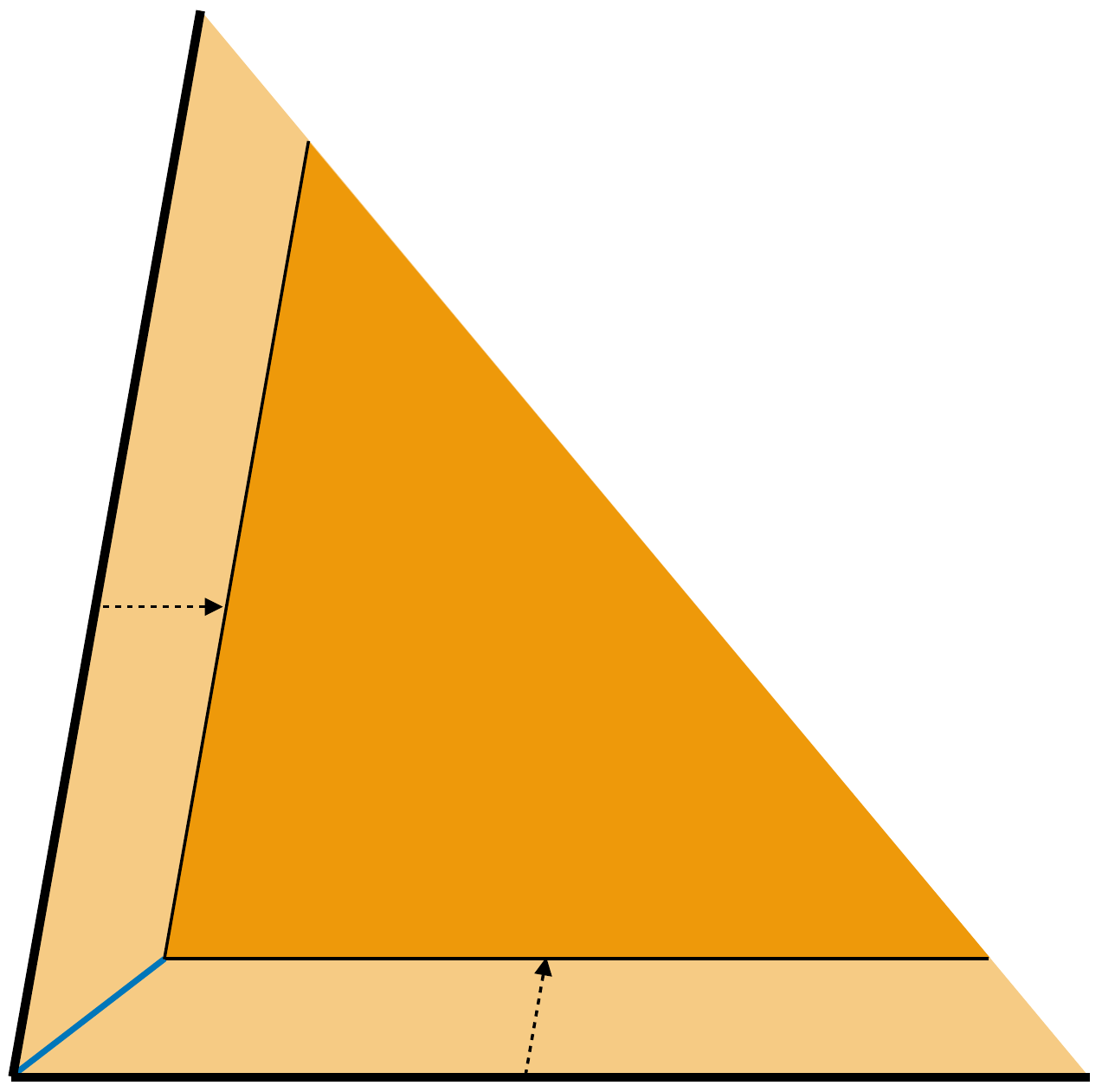}%
\begin{picture}(0,0)
\put(-74,7.5){\scriptsize$c$}
\put(-130,69){\scriptsize$c$}
\put(-130,6.5){\scriptsize$d_{\rm min}$}
\end{picture}
\caption{Wide K\"ahler cone.\label{fig_01_a}}
\end{subfigure}
\hspace{70pt}
\begin{subfigure}{0.35\textwidth}
\centering
\includegraphics[width=150pt]{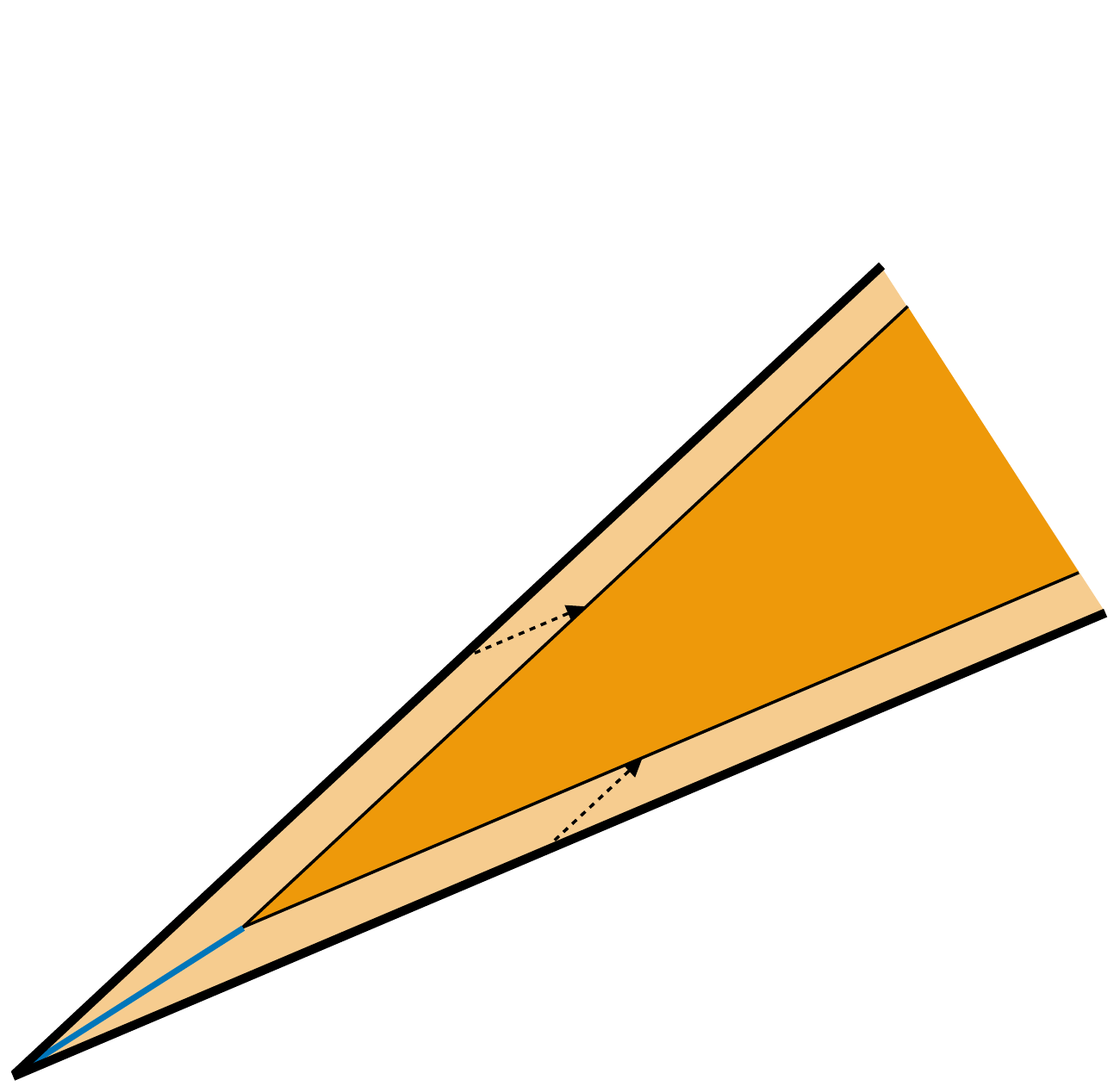}%
\begin{picture}(0,0)
\put(-90,83){\scriptsize$c$}
\put(-86,80){\vector(1,-2){6}}
\put(-52,19){\scriptsize$c$}
\put(-54,24){\vector(-1,1){10}}
\put(-102,6.5){\scriptsize$d_{\rm min}$}
\put(-105,10){\vector(-4,1){20}}
\end{picture}
\caption{Narrow K\"ahler cone.\label{fig_01_b}}
\end{subfigure}
\caption{K\"ahler and stretched K\"ahler cones $\mathcal K$ and $\widetilde{\mathcal K}$ 
for a two-dimensional setting. In figure~\ref{fig_01_a}
a wide cone is shown, while in figure~\ref{fig_01_b} we have displayed a narrow K\"ahler cone. 
The darker shaded regions are the stretched K\"ahler cones with parameter $c$,
and we have included the minimal distance $d_{\rm min}$ between the tip of the cone and the stretched cone 
 for both cases.
\label{fig_cones}}
\end{figure}

For compactifications of string theory on Calabi-Yau three-folds $\mathcal Y$,
the K\"ahler moduli space is bounded by the K\"ahler cone $\mathcal K_{\mathcal Y}$. 
This cone is defined as the dual of the Mori cone, and from a practical
point of view
it contains all K\"ahler forms $J$ such that curves and divisors in $\mathcal Y$
as well as the 
Calabi-Yau three-fold itself have positive volumes
\eq{
  \mathcal K_{\mathcal Y} = \left\{ 
  J \in H^{1,1}(\mathcal Y,\mathbb R) : 
  \textrm{vol}(W) >0 \quad\forall \,W \in \mathcal W\right\} ,
}
where $\mathcal W$ are all subvarieties (curves, divisors, $\mathcal Y$) of 
the compact space. 
This ensures in particular that the K\"ahler metric and the Hodge-star matrix are positive-definite. 
Since near a boundary in moduli space the effective theory 
typically receives corrections, it is useful to define 
a stretched K\"ahler cone $\widetilde{\mathcal K}_{\mathcal Y}$
such that all $W\in\mathcal W$ have a volume greater or equal 
to a constant $c$ (in appropriate units)
\eq{
  \label{skcone}
  \widetilde{\mathcal K}_{\mathcal Y}[\op c\op] = \left\{ 
  J \in H^{1,1}(\mathcal Y,\mathbb R) : 
  \textrm{vol}(W) \geq c \quad\forall \,W \in \mathcal W\right\} \,.
}
As illustrated in figures~\ref{fig_cones}, this means that inside the stretched K\"ahler cone 
there is a minimal distance to the boundary 
of moduli space.
In practice it can be computationally challenging to determine
the precise form of the K\"ahler cone, and therefore in \cite{Demirtas:2018akl} 
(see also \cite{Cicoli:2018tcq})
the authors 
considered a cone $\mathcal K_{\cap}$ for which $\mathcal W$ does not contain 
all subvarieties.  However, $\mathcal K_{\cap}$ contains the
$\mathcal K_{\mathcal Y}$ and one finds
\eq{
 \mathcal K_{\mathcal Y}\subseteq  \mathcal K_{\cap} \,,
 \hspace{70pt}
 \widetilde{\mathcal K}_{\mathcal Y}[\op c\op] \subseteq  \widetilde{\mathcal K}_{\cap}[\op c\op] \,.
}

\subsubsection*{Numerical results}

Next, we recall some numerical results of \cite{Demirtas:2018akl}
obtained for the Kreuzer-Skarke database  \cite{Kreuzer:2000xy}. 
The authors
determined properties of the K\"ahler cone for Hodge numbers $h^{1,1}$ 
in the range $2\leq h^{1,1}\leq491$ of the following form:
\begin{itemize}

\item  The K\"ahler cone $\mathcal K_{\mathcal Y}$ 
is bounded by hypersurfaces which intersect
the origin of moduli space. If the angle between these
hypersurfaces is small, the K\"ahler cone is narrow. 
In figure~3 of \cite{Demirtas:2018akl} the dependence of the (cosine of 
the) smallest angle on $h^{1,1}$ is shown, and one sees that for large $h^{1,1}$ 
this angle tends to become small. Hence, for larger $h^{1,1}$
the K\"ahler cone becomes narrower.

\item Another way to quantify the narrowness of the K\"ahler cone 
is to consider the stretched K\"ahler cone \eqref{skcone}.
This cone does not reach the origin of moduli space, and
one can determine the point closest to it. 
The narrower the K\"ahler cone the further away 
the minimal point, as illustrated in figures~\ref{fig_cones}.
More concretely, in \cite{Demirtas:2018akl}
the authors consider the stretched K\"ahler cone 
$\widetilde{\mathcal K}_{\cap}[c ]$ and 
determine 
the minimal distance 
$d^{\cap}_{\rm min}[c]$ from the origin.
This dependence on the Hodge numbers $h^{1,1}$ for $c=1$ 
is shown in 
figure~5 of \cite{Demirtas:2018akl}
and has been fitted as
\eq{
  d^{\cap}_{\rm min}[1] \simeq 10^{-1.4} \op (h^{1,1})^{2.5}  \,.
}
Note that this data has been determined for $c=1$, 
however, $d^{\cap}_{\rm min}$ depends linearly on $c$.
This implies we can generalize the above fit as
\eq{
  \label{cone_fit_01}
 d^{\cap}_{\rm min}[c] \simeq 10^{-1.4} \op (h^{1,1})^{2.5} \, c\,.
}

\item Finally, from  figure~5 in \cite{Demirtas:2018akl} we can also determine an approximate lower bound
in the region $h^{1,1}\gtrsim 10$ of the form 
\eq{
d^{\cap}_{\rm min}[c] \gtrsim 10^{-2.6} \op (h^{1,1})^{2.7}\, c  \,.
}

\end{itemize}


\subsubsection*{Mirror symmetry}

The complex-structure moduli space in the large complex-structure limit of a Calabi-Yau three-fold $\mathcal X$ 
is related to the K\"ahler moduli space of a Calabi-Yau manifold 
$\mathcal Y$ via mirror symmetry. 
For type IIB orientifolds discussed in this work the imaginary part of 
complex-structure moduli $z^i$ has to be inside a cone, which is the
K\"ahler cone on a mirror three-fold. 
We refer to this cone as the complex-structure cone.
From the results reviewed above, we 
can conclude that for large $h^{2,1}$ this cone becomes narrow. 
In particular, using the fit shown in \eqref{cone_fit_01} we can estimate the 
scaling of the minimal distance of the stretched cone to the origin as
\eq{
\label{tad_822}
 d^{\rm cs}_{\rm min}[c] \simeq 10^{-1.4} \op (h^{2,1})^{2.5} \, c\,,
}
where $c$ is the minimal distance away from the boundary of the ordinary cone. 
Similarly, the minimal distance away from the origin in a stretched K\"ahler cone 
is expected to satisfy the bound
\eq{
\label{tad_823}
  d^{\rm cs}_{\rm min}\gtrsim 10^{-2.6} \op (h^{2,1})^{2.7}\, c\,.
}


\subsection{Implications for the tadpole conjecture}
\label{sec_scaling}

In section~\ref{sec_tad_bound} we have argued that  when approaching a boundary in 
moduli space, the flux number $N_{\rm flux}$ typically diverges.
This applies in particular to a boundary of the complex-structure cone, and  for obtaining a minimal 
$N_{\rm flux}$ one should therefore stabilize moduli in the interior of moduli space. 
However, in section~\ref{sec_prop_cone} we have seen that for increasing 
$h^{2,1}$ the complex-structure cone becomes narrower and 
it  becomes more difficult to avoid the boundary regions.
Hence, for larger $h^{2,1}$ we expect an increase in $N_{\rm flux}$.


\subsubsection*{Setting}

We now want to make this heuristic argument more concrete. 
Our main technical assumption for the following is that when the 
complex-structure cone becomes narrower, the increase of $N_{\rm flux}$ 
along a direction towards a boundary of the complex-structure cone is faster than
along the large complex-structure direction. 
This implies that for larger $h^{2,1}$ one is pushed towards large 
complex-structure, when searching for a minimal flux number $N_{\rm flux}$.
Or, in other words, we assume that the lowest value for $N_{\rm flux}$ is found 
inside a stretched complex-structure cone with parameter $c$ that is not too small.


\subsubsection*{Lower bound on $N_{\rm flux}$}

Now, in equation \eqref{scaling_838} we have estimated that 
in the large complex-structure limit and for large $h^{2,1}$ 
the flux number for generic flux-compactifications behaves as
\eq{
  \label{tad_514}
 N_{\rm flux} \gtrsim (h^{2,1})^{1-\frac{\gamma}{2}} \op \lVert v \rVert^{\gamma} \,, \hspace{70pt}
 \gamma=1,3\,,
}
where $\lVert v \rVert$ denotes the Euclidean norm of $v^i = \mbox{Im}\, z^i$. 
We have furthermore argued that for large $h^{2,1}$ the complex-structure 
cone becomes narrow, and we determined in \eqref{tad_823} a lower bound on the minimal distance 
of a stretched K\"ahler cone to the origin of moduli space.
To obtain a minimal $N_{\rm flux}$ we therefore identify  $\lVert v \rVert$ in \eqref{tad_514} 
with $ d^{\rm cs}_{\rm min}$
and obtain the lower bound
\eq{
\label{tad_bound_004}
 N_{\rm flux}[c] \gtrsim (h^{2,1})^{1-\frac{\gamma}{2}}  \op \left[ 
10^{-2.6} \op (h^{2,1})^{2.7}\, c\right]^{\gamma} \,,
\hspace{50pt}
 \gamma=1,3\,.
}
This equation  depends on 
the parameter $c$ which parametrizes the minimal distance from
the stretched complex-structure cone to the boundary of 
the ordinary complex-structure cone (see figures~\ref{fig_cones}).
In particular, $c$ is the minimal volume of two-cycles 
on the mirror-dual side -- and for having control over the world-sheet instanton 
corrections one expects 
that $c$ should not be much smaller than one. 
In view of our assumption stated above, we also expect that for small $c$ the effect 
of being close to a boundary 
leads to an increase in the flux number $N_{\rm flux}$.


\subsubsection*{Comparison with the tadpole conjecture}

\begin{figure}[p]
\centering
\begin{subfigure}{0.9\textwidth}
\centering
\includegraphics[width=225pt]{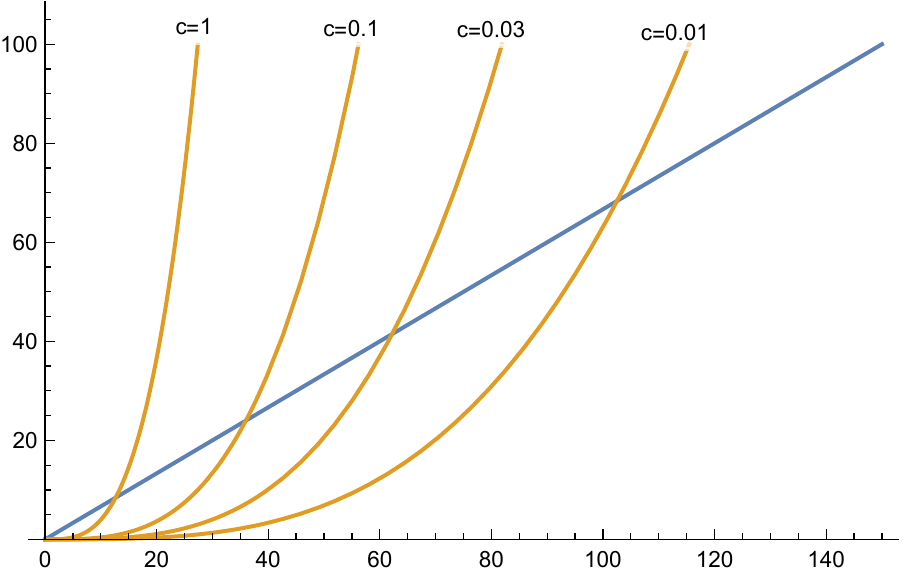}%
\begin{picture}(0,0)
\put(-238,140){\scriptsize$N_{\rm flux}$}
\put(2,2){\scriptsize$h^{2,1}$}
\end{picture}
\caption{Plot of $N_{\rm flux}$ for $\gamma=1$ and different values of $c$.\label{fig_02_a}}
\end{subfigure}
\\[25pt]
\begin{subfigure}{0.9\textwidth}
\centering
\includegraphics[width=225pt]{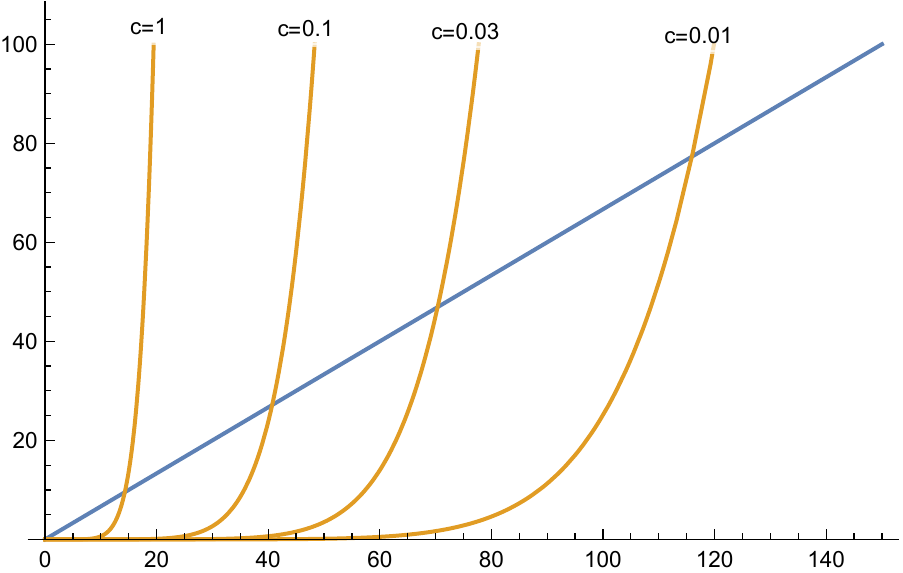}%
\begin{picture}(0,0)
\put(-238,140){\scriptsize$N_{\rm flux}$}
\put(2,2){\scriptsize$h^{2,1}$}
\end{picture}
\caption{Plot of $N_{\rm flux}$ for $\gamma=3$ and different values of $c$.\label{fig_02_b}}
\end{subfigure}
\caption{Comparison between the bound on $N_{\rm flux}$
shown in equation \eqref{tad_bound_004} (orange curves) 
and the bound from the tadpole conjecture \eqref{tad_conj} 
for $\alpha=1/3$ (blue curve).
In figure~\ref{fig_02_a} the parameter $\gamma$ is chosen as $\gamma=1$, while in 
figure~\ref{fig_02_b} we have chosen $\gamma=3$. 
\label{fig_nflux}}
\end{figure}
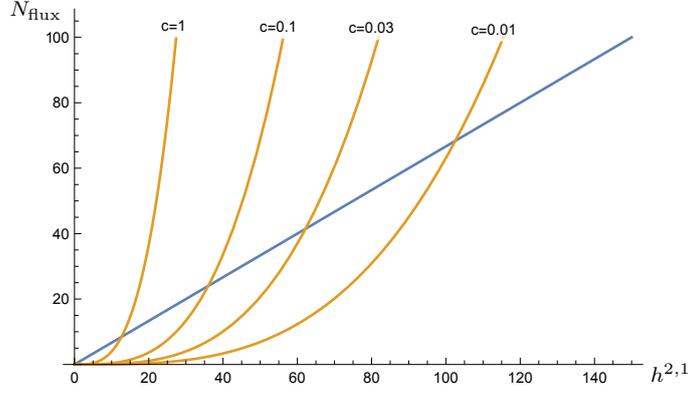
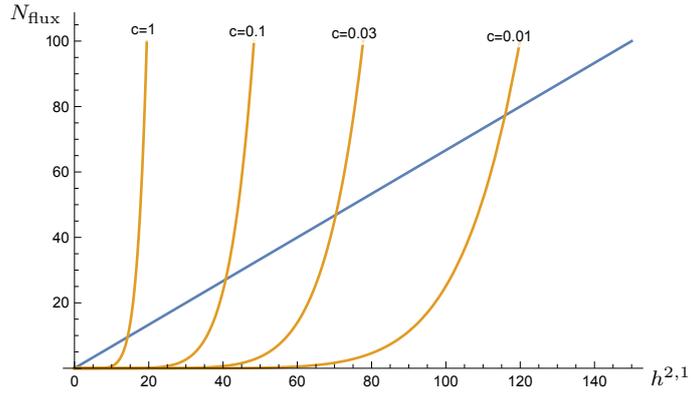

Within the approach we are following in this note we are not able to determine
a precise value of $c$.
We can however  compare the behavior of $N_{\rm flux}[c]$ for different 
values of $c$ with the  bound \eqref{tad_conj} of the tadpole 
conjecture.
We have shown this comparison in figures~\ref{fig_nflux},
from which we see that for large $h^{2,1}$ 
the bound  \eqref{tad_bound_004}  exceeds the linear bound of the 
tadpole conjecture. For larger values of $c$ this crossover happens
earlier, for smaller values of $c$ this happens later. 
In the large complex-structure and large $h^{2,1}$ regime,
for moderate values of $c$ and
for generic flux compactifications
the tadpole conjecture is therefore satisfied.

\subsubsection*{Summary and remarks}

We close this section with the following summary and remarks:
\begin{itemize}

\item We have estimated a  bound on how the flux number 
$N_{\rm flux}$ depends on the dimension of the 
complex-structure moduli space  
in the large complex-structure 
and large $h^{2,1}$ regime.  
To do so, 
we first determined for generic situations the scaling of 
$N_{\rm flux}$ with  $\lVert  v \rVert$
 which is shown in equation \eqref{scaling_838}.
Next, we have reviewed results of \cite{Demirtas:2018akl} 
on the geometry of the mirror-dual K\"ahler cone and how for 
large $h^{2,1}$ the corresponding complex-structure cone becomes narrow. 
Combining these observations we determined a lower bound on how  $N_{\rm flux}$ depends
on $h^{2,1}$, shown in equation \eqref{tad_bound_004}.

\item The bound in \eqref{tad_bound_004} depends on a parameter $c$, 
which parametrizes the minimal distance away from the 
boundary of the complex-structure cone. 
In order to  trust the large complex-structure approximation, this parameter
should not be  taken  much smaller than one. 
However, as shown in figures~\ref{fig_nflux}, even for values $c=0.1$ or $c=0.03$
we see that  the flux number $N_{\rm flux}$ 
exceeds the linear behavior \eqref{tad_conj} of the tadpole conjecture for moderately-large values of $h^{2,1}$. 
We have then concluded that in the large complex-structure regime, 
the tadpole conjecture is satisfied for generic Calabi-Yau three-folds.

\item We emphasize that the data in \cite{Demirtas:2018akl} 
is obtained for the Kreuzer-Skarke data\-base, which is expected to 
represent generic features of Calabi-Yau three-folds. 
Furthermore, when estimating the scaling \eqref{scaling_838} we have 
used an average value of $\mathsf v^i \simeq (h^{2,1})^{-1/2}$ for 
the components of the normalized vector $\mathsf v^i$.
We therefore have determined the generic behavior of $N_{\rm flux}$
in the large complex-structure and large $h^{2,1}$ regime,
which may be different for specific non-generic constructions.

\item Our main assumption in this section was that 
approaching a boundary of the complex-structure cone will lead to 
a faster increase of $N_{\rm flux}$ than 
approaching the large complex-structure point. 
In other words, when the complex-structure cone becomes narrower one is 
pushed towards the large complex-structure point. 
However, this assumption may  not be true.

\item Our analysis in this section has been 
carried out with the dilaton 
staying finite, in particular, $s$ does not scale with $h^{2,1}$. 
Such a scaling can be included, 
but for generic flux-configurations this will lead
to an even stronger growth of $N_{\rm flux}$. 
However, for specific and fine-tuned flux choices 
the growth of $N_{\rm flux}$ may be reduced. 
An example is $h^I=0$, $f^0\neq0$ and at least 
one $h_i\neq0$, together with the assumptions
that the K\"ahler metric $G_{i\ov j}$ is 
non-sparse and  
that the combinations $f^i-u^i\op f^0$ and $h_0+u^ih_i$ 
have a very specific dependence on 
$h^{2,1}$.
It is a non-trivial question whether such 
configurations can be realized in concrete examples, 
and we consider such constructions to be non-generic.

\end{itemize}


\clearpage
\section{Discussion}
\label{sec_disc}

Let us now summarize and discuss our arguments of the previous sections, and 
make some further remarks and comments on the tadpole conjecture. We start 
with a brief summary:
\begin{itemize}

\item In section~\ref{sec_tad_bound} we studied the boundary behavior of the 
flux number $N_{\rm flux}$. We argued that when approaching a boundary 
in complex-structure and axio-dilaton moduli space, this quantity typically diverges. 
We made this observation using a Bloch-Messiah decomposition 
of the Hodge-star matrix $\mathcal M$, however, the same 
result follows from asymptotic Hodge theory \cite{Schmid,CKS,Grimm:2020cda}.
This implies, that a flux-configuration with minimal $N_{\rm flux}$ will
stabilize moduli in the interior of moduli space, where one has typically
less computational control. 
Hence, it can become challenging to 
explicitly test the tadpole conjecture made in \cite{Bena:2020xrh}.

\item In section~\ref{sec_tad_cone} we considered the large complex-structure 
limit and estimated a lower bond on the scaling of $N_{\rm flux}$ with 
$h^{2,1}$ for large $h^{2,1}$ and for generic situations. We obtained this bound  using statistical data of 
\cite{Demirtas:2018akl}  on how the complex-structure cone becomes narrower 
for larger $h^{2,1}$. 
For plausible choices of a parameter $c$, 
we find that our bound exceeds the bound of the tadpole 
conjecture for moderate values of $h^{2,1}$ (see figures~\ref{fig_nflux}), and therefore 
we find support for the conjecture in the large complex-structure regime. 
However, a configuration with a flux number $N_{\rm flux}$ that violates 
the tadpole conjecture may 
be outside of this regime.

\end{itemize}
We finally want to add further comments and remarks in view of the tadpole conjecture 
 \cite{Bena:2020xrh}:
\begin{itemize}

\item We emphasize that the bound obtained in \eqref{tad_bound_004} is not 
a proof of the tadpole conjecture. 
But, independent of this conjecture, it shows that for large $h^{2,1}$ and for generic flux-configurations, 
stabilizing all complex-structure and axio-dilaton moduli 
by fluxes in the large complex-structure regime in a consistent way is rather difficult.

\item Even if isolated counter examples to the tadpole conjecture exist, 
the conjecture can have important implications for the size of the string-theory landscape. In particular, if the majority of 
string-compactifications does not allow for the  stabilization of sufficiently many moduli, 
the landscape of effective four-dimensional theories with no or few massless scalar fields
is smaller than naively expected. 

\item The constant $\alpha$  in the tadpole conjecture
\eqref{tad_conj} has been proposed as $\alpha=1/3$, however,
in \cite{Bena:2020xrh} it is mentioned that the conjecture is often 
also satisfied for $\alpha\simeq0.44$. 
We want to point out that  deriving an exact numerical value for $\alpha$ appears to be difficult,
since it emergences only at large $h^{2,1}$. Indeed, for 
$h^{2,1}=1$ one can find examples which stabilize the complex-structure modulus and the axio-dilaton 
with a flux number $N_{\rm flux}=1$, which would correspond to a value of $\alpha=1/4$. 
(We did this analysis for the conifold point 
of $X_{4,2}(1,1,1,1,1,1)$ which was discussed in \cite{Joshi:2019nzi}.)

\end{itemize}


\vskip3em
\subsubsection*{Acknowledgments}

We thank
Mariana Gra\~na,
Thomas Grimm,
Damian van de Heisteeg
and
Viraf Mehta
for very helpful discussions and communications. 
The work of EP is supported by a Heisenberg grant of the
\textit{Deutsche Forschungsgemeinschaft} (DFG, German Research Foundation) 
with project-number 430285316.


\clearpage
\nocite{*}
\bibliography{references}
\bibliographystyle{utphys}


\end{document}